# Nonlinear photodetector based on InSe *p-n* homojunction for improving spatial imaging resolution

*Yu Zhang*[1], *Xiaoqing Chen*[1,*], *Mingwen Zhang*[1], *Xianghu Wu*[1], *Jianguo Wang*[1], *Ruijuan Tian*[1], *Liang Fang*[1], *Yanyan Zhang*[2], *Jianlin Zhao*[1], *Xuetao Gan*[1,3,*]

[1]Key Laboratory of Light Field Manipulation and Information Acquisition, Ministry of Industry and Information Technology, and Shaanxi Key Laboratory of Optical Information Technology, School of Physical Science and Technology, Northwestern Polytechnical University, Xi'an, 710129 China

[2]School of Artificial Intelligence, OPtics and ElectroNics (iOPEN), Northwestern Polytechnical University, Xi'an, 710072, China

[3]School of Microelectronics, Northwestern Polytechnical University, Xi'an, 710129, China

E-mail: xuetaogan@nwpu.edu.cn, xiaoqing_chen@nwpu.edu.cn

**Abstract**

We demonstrate an efficient nonlinear photodetector (NLPD) with quadratic response based on a few-layer InSe *p-n* homojunction, which is beneficial from the strong second harmonic generation (SHG) process in InSe and effective harvest of photocarriers actuated by the high-quality homojunction. The NLPD can sense light with photon energy smaller than InSe's electronic bandgap because the SHG process in InSe doubles the frequency of incident light, extending InSe's photodetection wavelength range to 1750 nm. The InSe *p-n* homojunction, which is electrostatically doped by two split back gates, presents a rectification ratio exceeding $10^6$ with a dark current down to 2 pA and a high normalized responsivity of 0.534 A/W$^2$ for the telecom-band pulsed light at 1550 nm. The photocurrents of the SHG-assisted photodetection have a quadratic dependence on the optical powers, making the NLPD highly sensitive to light intensity variation with improved spatial resolution. As examples, the NLPD is employed to precisely determine the localization point of a focused laser beam waist and implement spatial imaging with an improved resolution compared with the linear photodetector. These features highlight the potential of the proposed NLPD in developing advanced optical sensing and imaging systems.

## 1. Introduction

Nonlinear photodetectors (NLPDs), which have nonlinear dependences between the measured photo-currents or -voltages and the optical powers, can perform sophisticated optoelectronic functions, including direct photoelectric conversion of all-optically mixing signals or in-sensor computing of photoelectric conversion signals.[1-4] It can simplify complicated optoelectronic systems. For instance, in the autocorrelation measurement of ultrafast laser pulse width, the optical pulse-pulse interaction into frequency conversion signals is carried out first using a nonlinear crystal, which is then converted into an electrical signal by a separated photodetector.[5] A NLPD based on second- or third-order nonlinear responses can realize this autocorrelator directly.[6] In telecommunications, to mix two optical signals modulated in radio frequency into an electrical signal, an optoelectronic mixing system based on two photodetectors and an electrical mixer is required,[7] which actually could be realized by a single NLPD with all-optical mixing in-sensor.[2]

NLPDs have been demonstrated based on the mechanisms of multi-photon absorption,[8] saturable absorption,[9] carrier-defect trapping.[10] For the NLPDs based on saturable absorption or carrier-defect trapping, because various physical processes are involved in the complex interplay, it is challenging to determine the function between the responsivities and the optical powers. The NLPDs governed by multi-photon absorption face limitations in photoresponsivity due to inefficient higher-order nonlinear optical responses. Recently, we reported a NLPD by van der Waals (vdWs) stacking a *p-n* heterojunction of few-layer gallium selenide (GaSe) and few-layer indium selenide (InSe).[11] Benefiting from the strong second-harmonic generations (SHGs) in GaSe and InSe, the heterojunction upconverts two of the incident photons with frequency of $\omega$ into one photon with frequency of $2\omega$, which is then absorbed by the heterojunction to generate one electron-hole pair for the photocurrent. It results in a quadratic photoelectric response in this vdWs NLPD. Thanks to its capability for optical frequency upconversion as well as the direct photoelectric conversion of light-light interaction, this NLPD not only extends the photodetecting range of GaSe/InSe from 900 nm to 1750 nm, but also supports the functions of an autocorrelator for measuring ultrafast pulse widths and an optoelectronic mixer of two modulated pulses for signal processings.

Here, we improve the SHG-enabled NLPD by replacing the GaSe/InSe vdWs heterojunction with an InSe *p-n* homojunction, as schematically displayed in **Figure 1a**. A few-layer InSe contacted by drain and source electrodes (D, S) is back-gated by two split metal electrodes (G1, G2), where a few-layer hexagonal boron nitride (h-BN) is employed as the dielectric layer. Applying gate voltages with varied signs and magnitudes to G1 and G2 enables doping of the InSe channel into *p* or *n*-type over the respective gate regions, forming *p-n* homojunctions. These junctions provide the necessary built-in electric field for separations of photocarriers. This homojunction-based photodetector could operate in the mechanisms of linear photoelectric conversion (LPC) and nonlinear photoelectric conversion (NPC) for light with photon energy larger and smaller than InSe's bandgap respectively, as schematically shown in Figure 1b. Combining with the capability of strong SHG process in the InSe,[12] the photodetector with NPC performs as a NLPD with quadratic photoelectric response for the light with photon energy smaller than the bandgap of InSe, which is undetectable by the linear optical absorption.

Compared with our previous NLPD based on GaSe/InSe vdWs heterojunction, the InSe homojunction offers natural matched chemical and electronic structures across the electrical channel, reducing interface defects and dislocations density, and promising an enhanced carrier diffusion channel with continuous energy band bending.[13] This advancement leads to a significant increase in photoresponsivity, reaching 227 μA/W at an averaged optical power of 425 μW under 1550 nm pulsed laser, over two orders of magnitude higher than that obtained in our previous NLPD of GaSe/InSe vdWs heterojunction.

Furthermore, a constant quadratic relationship between the incident optical power ($P_{\text{laser}}$) and the photocurrent ($I_{\text{ph}}$) is verified in this InSe homojunction NLPD, specifically $I_{\text{ph}} \propto P_{\text{laser}}^2$. This feature is crucial for high-localization measurements of a focused beam waist and improving the spatial resolution in patterned imaging, offering superior performances over the linear photodetector, as schematically shown in Figure 1c and 1d. The *p-n* homojunction is defined by two split gates, leaving a narrow line of the built-in field to separate the photocarriers. Hence, the NLPD has a one-dimensional active profile with a width in the micrometer scale, which can be used to resolve the distribution of an optical beam by scanning across it. The improved resolution can be attributed to the heightened sensitivity to optical intensity in the NLPD governed by $I_{\text{ph}} \propto P_{\text{laser}}^2$, whereas in the LPC regime it has the relation of $I_{\text{ph}} \propto P_{\text{laser}}$. These features underscore the NLPD's versatility in delivering advanced optical sensing and imaging capabilities.

## 2. Results and discussion

**Figure 2a** shows the optical microscope image of the fabricated device. The fabrication of the device is accomplished by sequentially dry-transferring and vdWs stacking h-BN, InSe, D and S electrodes onto the pre-deposited G1 and G2 electrodes, as detailed in the section of *Device Fabrication*. To ensure efficient SHG, a few-layer InSe with a thickness of approximately 52 nm is utilized, as confirmed by the atomic-force microscope. The 48 nm h-BN dielectric layer ensures a clean vdWs interface with the InSe channel, preventing unexpected doping by the substrate. At the bottom of the device, the two gold (Au) gate electrodes (G1 and G2) are defined using electron beam lithography to have a lateral gap of 150 nm. The D and S electrodes are dry-transferred onto the InSe channel to form their vdWs contacts, ensuring high-quality electrical connectivity.[14] Here, Au is chosen as the D and S electrodes because its work function locates close to the middle of InSe's bandgap, facilitating the injections of holes and electrons into InSe which leads to the ambipolarity.[15] Raman spectroscopy and photoluminescence (PL) pumped under 532 nm excitation confirm the InSe's quality. The Raman spectrum presents peaks at 115, 177 and 227 cm$^{-1}$ (see Figure S1a in the Supporting Information), corresponding to the $A_{1g}^1$, $E_{2g}^1$, and $A_{1g}^2$ vibrational modes of ε-InSe, indicative of its high crystallinity.[12,16] The PL spectrum indicates the bandgap of the ε-InSe in this work is about 1.26 eV (see Figure S1b in the Supporting Information), which is consistent with the previous reports.[12,17]

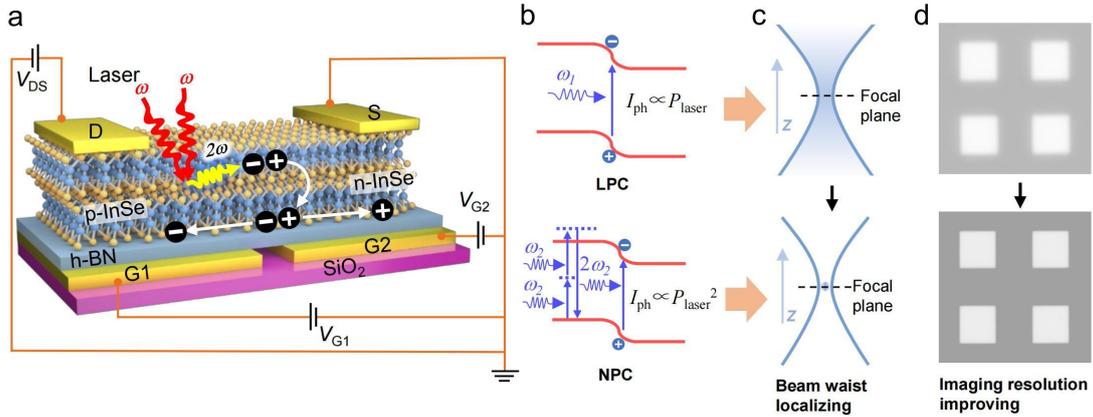

**Figure 1.** Device concept, photoelectric conversion mechanisms and applications of the proposed NLPD. **a** Schematic of the NLPD based on InSe homojunction tuned by two split back gates with h-BN as the dielectric layer, showing the combined processes of SHG and separation of photocarriers. **b** Mechanisms of linear photoelectric conversion (LPC, upper) and nonlinear photoelectric conversion (NPC, lower) in the homojunction-based photodetector. For the LPC, the incident photons (ω$_1$) with the energy larger than the InSe bandgap are directly absorbed to produce electron-hole pairs, leading to $I_{ph} \propto P_{laser}$; For the NPC, two photons (2*ω$_2$) with the energy smaller than the InSe bandgap undergo frequency doubling to produce one photon (2ω$_2$) with the energy greater than the InSe bandgap and then is absorbed to produce electron-hole pairs, leading to $I_{ph} \propto P_{laser}^2$. **c,d** Two applications based on different photoelectric conversion mechanisms (from LPC to NPC). **c** Precise determination of the localization point of a focused beam waist by the NPC. **d** Improved optical imaging resolution by the NPC.

In exploring the dual-gated InSe *p-n* homojunction for the proposed NLPD, we first examine the electron-hole bipolar property of the InSe channel. By synchronizing the two split gates (G1 and G2) to a uniform gate voltage ($V_G$) and gradually varying $V_G$ from -8 V to 8 V, the measured transfer curve is obtained with the drain-source voltage ($V_{DS}$) as 3.5 V, as depicted in Figure 2b. Notably, the channel current ($I_{DS}$) reaches a minimum at $V_G$ = 0 V, indicating a balance in the injection of electrons and holes because there is an equivalent effect on the Fermi level's shift towards the conduction and valence bands with the work function of the Au electrode locating in the middle of the InSe bandgap.[15,18] For $V_G$ > 0 V and $V_G$ < 0 V, $I_{DS}$ increases significantly for the enlarged magnitude of $V_G$, showing an on-off ratio exceeding 10$^3$. It indicates the gate voltages effectively modulate the injections of both electrons and holes, thereby controlling the operational states of the InSe channel.

As a consequence, when opposite voltages are applied to the two split gates, the corresponding regions of the InSe channel located above the gates could be doped into *n*- or *p*-type, giving rise to a *p-n* homojunction around the region over the gap between the two gate electrodes.[19] We depict the band diagrams of various two gates configurations under $V_{DS}$ > 0, illustrating the carrier injection dynamics (Figure 2c). When positive voltages are applied to both $V_{G1}$ and $V_{G2}$, the conduction and valence bands are downward bent (Figure 2 (c-I)). Electrons can be injected at the source terminal due to the thinned Schottky barriers, while the increased Schottky barriers at the drain terminal blocks the hole injection, leading to *n*-type doping of the InSe channel. Conversely, negative voltages applied on both $V_{G1}$ and $V_{G2}$ induce upward bending of the conduction and valence bands (Figure 2 (c-II)), which prohibits the injection of electrons through the source terminal while holes are injected at the drain terminal, thus inducing *p*-type doping of the InSe channel. When $V_{G1}$ < 0 and $V_{G2}$ > 0 (Figure 2 (c-III)), the energy bands at the source and drain terminals are bent in opposite directions, and the thinned Schottky barriers allow the injection of electrons and holes respectively. Under the configuration of $V_{DS}$ > 0, the device is conductive for the above three configurations, which corresponds to three distinct on-current states. When $V_{G1} > 0$ and $V_{G2} < 0$, the device exhibits off-current state due to the reverse bias of the built-in potential in the depletion region of the n-p junction (Figure 2 (c-IV)). Furthermore, by varying the direction of $V_{DS}$ to align with or oppose to the built-in electric field direction in the *p-n* junction, the current switches between the on and off states.

To reveal the diverse electrical characteristics of the fabricated NLPD device, the output characteristics are measured under different configurations of two gates voltages, as shown in Figure 2d. Four different channel dopings types of *n-n*, *p-p*, *p-n* and *n-p* are achieved. In the *n-n* ($V_{G1} = V_{G2} = 8V$) and *p-p* ($V_{G1} = V_{G2} = -8V$) configurations, the output curves are basically symmetric where electrons or holes in the channels act as the predominant carriers to determine the conductance of the device. However, distinctly asymmetric behaviors are observed in the *p-n* ($V_{G1} = -8V$, $V_{G2} = 8V$) and *n-p* ($V_{G1} = 8V$, $V_{G2} = -8V$) configurations. This is attributed to the injection of electrons and holes through the source and drain terminals, respectively, under positive $V_{DS}$ bias (i.e., $V_{DS} > 0$), resulting in forward conduction behavior in the *p-n* type channel. However, due to the blocking of electron and hole injection with $V_{DS} > 0$, the conductivity of the *n-p* type channel is suppressed. Moreover, the symmetric outputs of *n-n* and *p-p* configurations imply the symmetric contacts and the clean channel without residues, and therefore the rectifications in *n-p* and *p-n* configurations are principally originated from the *p-n* homojunction rather than Schottky barriers at the Au/InSe contacts.[20] Regarding optoelectronic devices with low static power consumption and a high on/off ratio are preferred,[21] our results (see Figure S2 in the Supporting Information) reveal that the InSe homojunction exhibits an off-current level below $10^{-11}$A and demonstrates on/off ratios of $10^5$ and $10^6$ for *p-n* and *n-p* junctions, respectively.

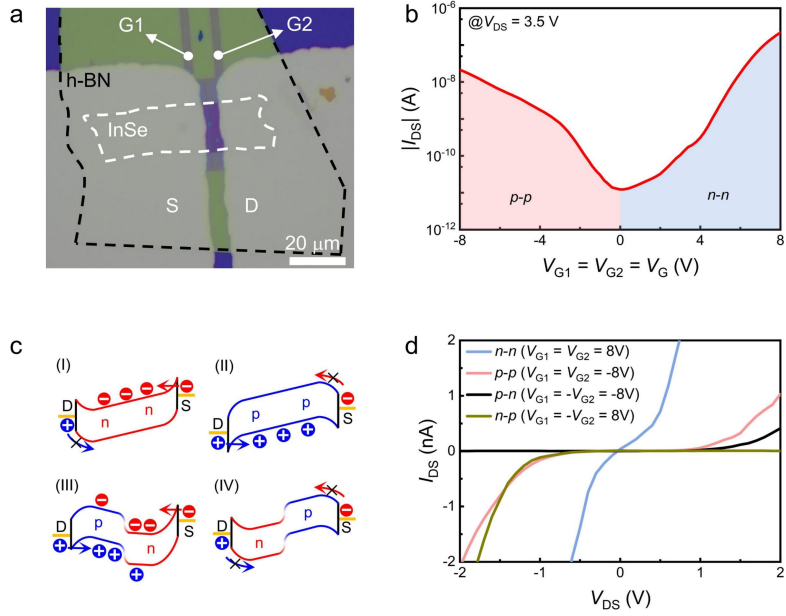

**Figure 2.** Electrostatic control of the InSe homojunction. **a** Optical micrograph of the InSe homojunction device. **b** Transfer characteristic ($|I_{DS}|$-$V_G$ curves) at $V_{DS}$ = 3.5 V with a minimum channel current $I_{DS}$ at $V_G$ = 0 V. **c** Band diagrams for different types of homojunctions under $V_{DS}$ > 0. **d** Output characteristics ($I_{DS}$-$V_{DS}$ curves) of four doped configurations of the device: *n-n*, *p-p*, *p-n* and *n-p* homojunctions.

The above high-quality InSe *p-n* homojunction promises the effective separation of photocarriers for sensitive photodetections. We evaluate it by illuminating the device with a continuous wave (CW) laser at the wavelength of 532 nm, which is focused using an objective lens (50×, numerical aperture of 0.75). Since the photon energy of 532 nm (2.3 eV) is larger than the bandgap of InSe (1.26 eV), photons can be directly absorbed by the InSe homojunction and lead to considerable photocurrents, as shown in **Figure 3a**. Here, to maintain the configuration of *p-n* homojunction, the back gates of $V_{G1}$ = -8V and $V_{G2}$ = 8V are applied. In the dark condition, the drain-source current $|I_{DS}|$ is low as $10^{-12}$ A with the $V_{DS}$ = -1V. With the 532 nm light illumination, $|I_{DS}|$ increases remarkably with the increased optical powers. By comparing $|I_{DS}|$ in dark and light illumination, we derive the photocurrent ($I_{ph}$) and calculate the photoresponsivity ($R = I_{ph}/P_{laser}$), achieving a high responsivity of 0.4 A/W. We also measure the power dependence of the photocurrent, as shown in Figure 3b, presenting a linear function, i.e., $I_{ph} \propto P_{laser}$. It confirms that the photoelectric response of the fabricated InSe photodetector is governed by the photovoltaic effect of the *p-n* junction, eliminating the possible photoelectric effect of photogating induced by carrier trappings.[22-23]

We then change the illumination light from the 532 nm CW laser into a pulsed laser at the wavelength of 1550 nm, which has a pulse width of 4 ps and a repetition rate of 100 MHz. The photon energy (0.8 eV) of the 1550 nm light is insufficient to make the InSe be absorptive due to its small bandgap. Thanks to the giant second-order nonlinear susceptibility of InSe,[12] it converts two photon of 1550 nm into a photon with the energy of 1.6 eV via the SHG process, which is then absorbed by the InSe to generate photocarriers, as schematically shown in Figure 1a. Subsequently, considerable photocurrents are realized driven by the built-in electric field of the *p-n* junction. Figure 3c displays the $|I_{DS}|$-$V_{DS}$ curves measured in dark and with the illumination of the 1550 nm pulsed laser. Similar to the photoelectric responses of the 532 nm CW laser shown in Figure 3a, $|I_{DS}|$ increases significantly with the increased optical powers, from $10^{-12}$ A to $10^{-7}$ A at the reverse bias. The photocurrents obtained at different optical powers are extracted from these curves, as shown in Figure 3d (red line), presenting a quadratic dependence with the optical powers as $I_{ph} \propto P_{laser}^2$, which therefore defines our NLPD with the quadratic photoelectric response. Different from the definition of photoresponsivity of linear photodetector, the NLPD has a varied photoresponsivity for different optical powers. Here, the responsivity up to 227 µA/W is achieved at the incident power of 425 µW, corresponding to a normalized value of 0.534 A/W$^2$. It is more than two orders of magnitude higher than that obtained in our previous work based on a GaSe/InSe vdWs heterojunction.[11] It is beneficial from the better transport of carriers across the continuous InSe channel comparing with the vdWs interfaces of GaSe and InSe. At the same time, this also surpass the responsivity of numerous NLPD utilizing alternative photoelectric conversion mechanisms.[24-27] We also evaluate the photoelectric response of the NLPD with a CW laser at the wavelength of 1550 nm. Due to the low SHG efficiency under the pump of CW laser, there is no observable photoelectric response (see Figure S3 in the Supporting Information).

To elucidate the photocurrent mechanism, we also capture the SHG radiation from the InSe channel using the objective lens. Figure 3d (blue line) displays the pump power dependence of the SHG signal, presenting a slope close to 2.0, indicating effective SHG photons conversion into the acquired photocurrents. The spatial mappings of the radiated SHG signal and the converted photocurrent with the illumination of the 1550 nm pulsed laser are carried out (see Figure S4 of the Supporting Information). A uniform distribution of SHG is obtained over the whole InSe channel. In contrast, the photocurrent predominantly distributes around the gap region between the two back gate electrodes, which is the location of the InSe *p-n* homojunction. It is consistent that the built-in electric field of the *p-n* homojunction separates the photocarriers generated by the absorption of the SHG signal. In addition, over the boundary regions of the D and S electrodes, there is no observable photocurrent though the radiated SHG signal is strong. It further suggests the negligible Schottky barriers at the metal contacts of D and S electrodes, and the considerable strong built-in electric filed provided by the *p-n* homojunction.

The wavelength dependence of the quadratic photoelectric response and the radiated SHG signals in the NLPD is then examined by tuning the wavelength of the pulsed laser, as shown in Figure 3e. Both the SHG intensity (blue line) and the corresponding quadratic photocurrents (red line) exhibit a discernible reduction as the laser wavelength increases from 1500 nm to 1750 nm. This diminishing trend is attributed to the less efficient SHG process at longer wavelengths.

The homojunction's photoelectric response, unhindered by carrier trappings or defect-induced photogating,[28-29] promises a fast operation speed. To examine that, we switch on and off the 1550 nm pulsed laser using a mechanical chopper, and the modulated photocurrents are recorded, as shown in Figure 3f. The rising and falling times are evaluated as about 80 μs, which are evaluated at the modulation levels ranging from 10% to 90% of the maximum photocurrent. Factors like contact resistance, channel resistance, and device area primarily constrain this response speed.[30-31] Optimizing these factors can further improve the response speed.

Additionally, we also extended the investigation of linear and quadratic photoelectric responses to other homojunction configurations with the same device, as detailed in Figure S5 of the Supporting Information. The measurement results in the *n-p* homojunction are consistent with those from the *p-n* homojunction. This concurrence emphasizes the robustness and reliability of the device performance across different junction configurations.

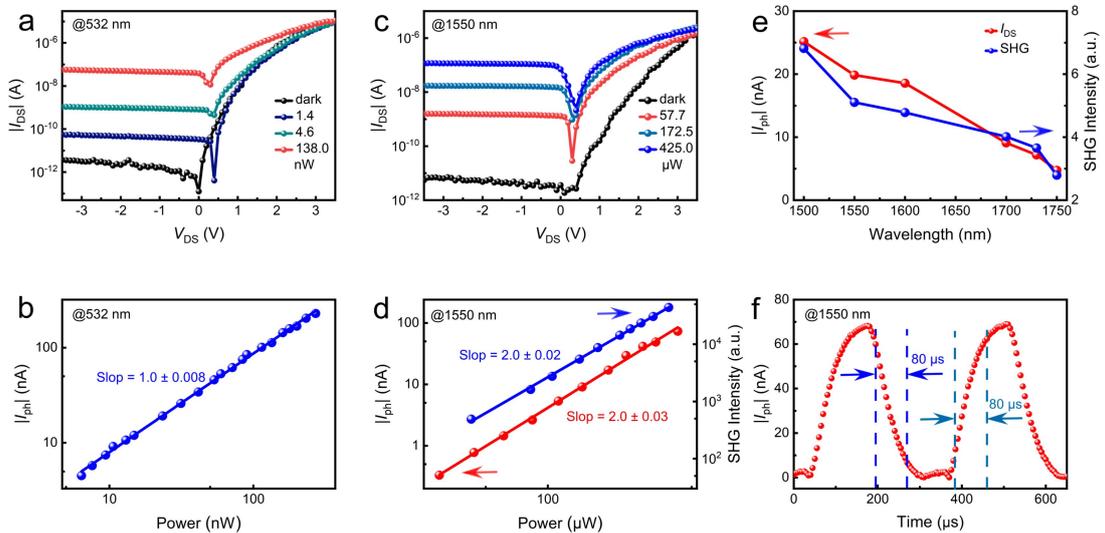

**Figure 3.** Linear and quadratic nonlinear photoelectric responses of the InSe *p-n* homojunction ($V_{G1}$ = -8V, $V_{G2}$ = 8V) photodetector. **a** |$I_{DS}$|-$V_{DS}$ curves for the illumination of a 532 nm CW laser with photon energy larger than InSe bandgap under different optical powers. **b** 532 nm CW laser power dependence of the photocurrent at $V_{DS}$ = -3.5 V, presenting a linear function, i.e., $I_{ph} \propto P_{laser}$. **c** |$I_{DS}$|-$V_{DS}$ curves for the illumination of a 1550 nm pulsed laser with photon energy smaller than InSe bandgap under different optical powers. **d** 1550 nm pulsed laser power dependence of the photocurrent (red line) at $V_{DS}$ = -3.5 V, presenting a quadratic function, i.e., $I_{ph} \propto P_{laser}^2$. The SHG signals (blue line) under different pulsed laser powers exhibit a slope of 2.0. **e** Variations of quadratic nonlinear photocurrent and SHG intensity with pump laser wavelength in the range of 1500 to 1750 nm. **f** Time-dependence of the quadratic nonlinear photocurrent under 1550 nm pulsed laser.

For the light with photon energy smaller than the bandgap of InSe, the photoelectric response of the NLPD presents the quadratic dependence of the photocurrent on the optical power, which strengthens the sensitivity to the variation of optical powers. Leveraging this attribute, we demonstrate two possible applications of NLPD in improving resolutions of the spatial measurements of optical signals. First, we show the quadratic photoelectric response offers a unique capability for achieving high-resolution measurements of the localization point of the focused laser beam waist. Accurate focused laser beam waist measurements not only facilitate the precise optimization of the focusing performance of laser systems but also provides a foundation for precision machining and optical design.[32-33] Common measurement methods including knife-edge measurement,[34] near-field measurement,[35] and micro/nano-particle scattering method[36] are all based on the linear optical response, which makes it difficult to determine the precise profiles and locations of the beam waist. We overcome this limitation using our NLPD. In our measurement, we employ an objective lens with a numerical aperture of 0.8 to focus the laser to form a beam waist. As schematically shown in **Figure 4a**, the NLPD is spatially scanned in the *x-z* plane around the focusing point. Since the photocurrents can only be generated when the laser is focused on the middle *p-n* junction region, the NLPD could be considered as a sensor with a one-dimensional geometry along the *y* direction. It is therefore analogous to a line photodetector, dynamically positioning itself at various focal depths (*z*) and along the focal plane (*x*). This dynamic positioning mirrors the concept of slicing the light profile with the line detector, enabling precise acquisition of measurement data pertaining to the focused beam waists.

To illustrate the advantages of the quadratic photoelectric response of the NLPD, we first use it to measure the focused beam waist of a 532 nm CW laser, which operates in the regime of linear photoelectric conversion response. The measurement result of the spatial scanning in *x-z* plane is shown in Figure 4b. By acquiring the photocurrents generated at different locations of the focused beam, the spatial profile as well as the intensity distribution of the focused laser beam is reconstructed. It could be well fitted by the focusing profile of a Gaussian beam. By changing the 532 nm CW laser into the 1550 nm pulsed laser, the focused beam waist is measured by the regime of nonlinear photoelectric response of the NLPD, as shown in Figure 4c. Different from that obtained with the linear photoelectric conversion response, the strengthened sensitivity by the quadratic dependence of the photocurrents on the optical powers ($I_{ph} \propto P_{laser}^2$) singularizes a highly localized focal point.

To provide a more comprehensive depiction of the enhanced localization of the focused beam under quadratic photoelectric response of the NLPD, we extract the acquired data curves at varying focal depths (*z*-direction) and along the focal plane (*x*-direction) under both linear and nonlinear photoelectric conversion responses. For the linear photodetection, the photocurrents along the *z*-direction have no significant peak, as shown in Figure 4d. The photocurrent remains consistently flat at a 5 μm depth of focus from the focal plane, and even at 10 μm from the focal plane, it experiences only a 12.2% reduction and maintains uniform across a broad range of depths. In contrast, as depicted in Figure 4e, the quadratic photodetection induces a distinct peak of the photocurrents at the focal plane, which decreases sharply out of the focal plane. The photocurrent diminishes to half of its maximum value at a depth of approximately 2.5 μm and approaches zero at around 8 μm from the focal plane. This phenomenon underscores the concentrated nature of the photocurrent under the quadratic photoelectric conversion response, which is primarily confined to the vicinity of the focal plane. The rapid decay of the photocurrent to zero at the defocusing plane is consistent with the high intensity solely at the focal plane, which promises the measurement accuracy and reliability of focused laser beam waist. The measured profiles along the *x*-direction at the focal plane are plotted in Figure 4f and 4g. We employ the full width at half maximum (FWHM) as a metric to quantify this distribution.[37] Generally, the FWHM of a focused beam waist is proportional to the laser wavelength, i.e., the laser at 1550 nm should have a wider FWHM than the laser at 532 nm. Intriguingly, the measured FWHM of the laser at 1550 nm with the quadratic photocurrent is about half of that measured from the 532 nm laser determined by the linear photocurrent data. This result is governed by the more sensitive of the nonlinear photoelectric conversion response in NLPD than the linear photoelectric response to the optical intensity. As a result, the NLPD based on the InSe *p-n* homojunction presents outstanding performance in measuring the location and profile of the highly focused beam waist, which has potentials to facilitate the examinations of laser beams in microscope, microfabrication, and imaging systems.

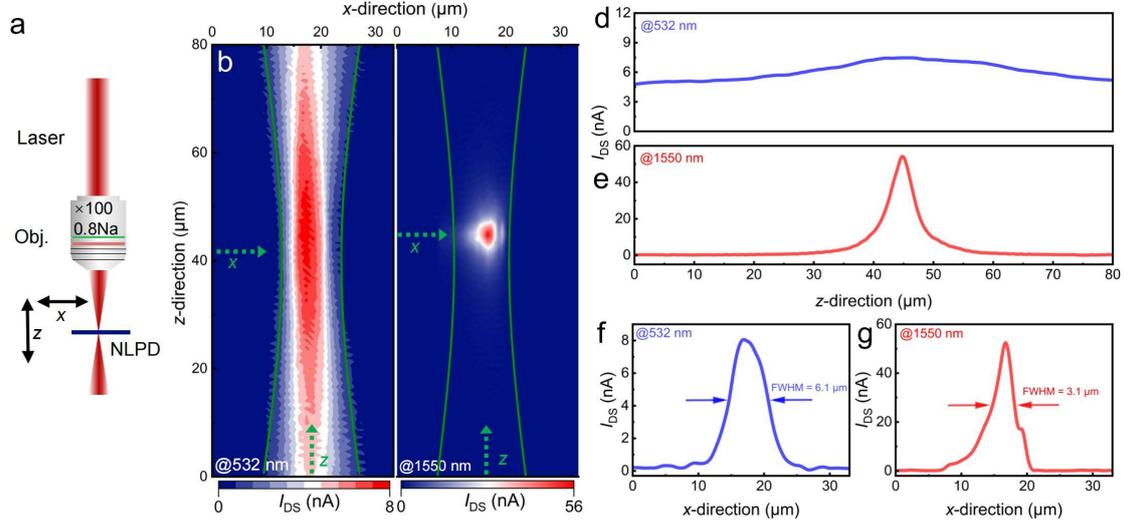

**Figure 4.** Measurement of the focused beam waist with the NLPD in linear and quadratic photoelectric conversion regimes. **a** Experimental setup of the focused beam waist measurement. **b** Measured profile of the focused beam waist of a 532 nm laser in the linear photoelectric conversion regimes. **c** Measured profile of the focused beam waist of a 1550 nm laser in the nonlinear photoelectric conversion regimes, displaying a highly localized point at the focal point. **d-e** Extracted measured photocurrent curves from (**b-c**) at varying focal depths (*z*-direction) under linear (**d**) and nonlinear (**e**) photoelectric responses. **f-g** Extracted measured photocurrent curves from (**b-c**) along the *x*-direction at the focal plane, indicating the FWHM of the focused beam waist under linear (**f**) and nonlinear (**g**) photoelectric responses.

The second application of the NLPD is the improvement of spatial imaging resolution relying on the quadratic dependence of the photocurrent on the optical power. To implement that, we home-build a confocal transmission laser scanning microscope setup, as schematically shown in **Figure 5a**. The imaging object is a mask with opaque metal square patterns deposited on a transparent glass substrate, which has squares with width of 3 μm and lattice spacing of 3 μm. To demonstrate the superiority of the quadratic photoelectric conversion response over the linear photoelectric conversion response in the imaging, lasers with the wavelengths of 1550 nm and 532 nm are chosen as illumination sources. The incident laser is focused by an objective lens (with a numerical aperture of 0.42) on the imaging mask, which is then collimated by another objective lens. The transmitted light is finally detected by the NLPD. By spatially scanning the mask in-plane at the focal point of the objective lens and acquiring the photocurrents in the NLPD, the opaque metal patterns could be imaged with a localization defined by the focused beam waist.

The comparative imaging results under linear and quadratic photoelectric conversion responses are presented in Figure 5b, where the green dashed lines indicate the location of the opaque metal squares. The imaging result based on the quadratic photoelectric conversion response exhibits clear profiles at the metal squares, while the counterpart based on linear photoelectric conversion response appears blurry and lacks well-defined contours. The FWHM of a grayscale curve extracted from the photocurrent mapping around the metal square serves as a quantitative metric to assess the imaging resolution,[38-39] as depicted in Figure 5c. The FWHM in the case of quadratic photoelectric conversion response is about 2.88 μm, which is narrower than that obtained with the linear photoelectric conversion response (3.26 μm). Michelson contrast is also a key metric for assessing image quality and is defined as $Contrast = (I_{max} − I_{min})/(I_{max} + I_{min})$,[40] where $I_{max}$ and $I_{min}$ represent the maximum and minimum intensity values of the grayscale curve, respectively. Calculations based on the data from Figure 5c reveal that the contrast for quadratic photoelectric conversion response image is about 3.35 times higher than that of linear photoelectric conversion response image. The notable increase in contrast means that the brightness differences between adjacent areas in the image are more pronounced, which not only reveals more image details but also visually enhances the perception of resolution. This underscores the superiority of quadratic photoelectric conversion response of the NLPD in improving the imaging resolution.

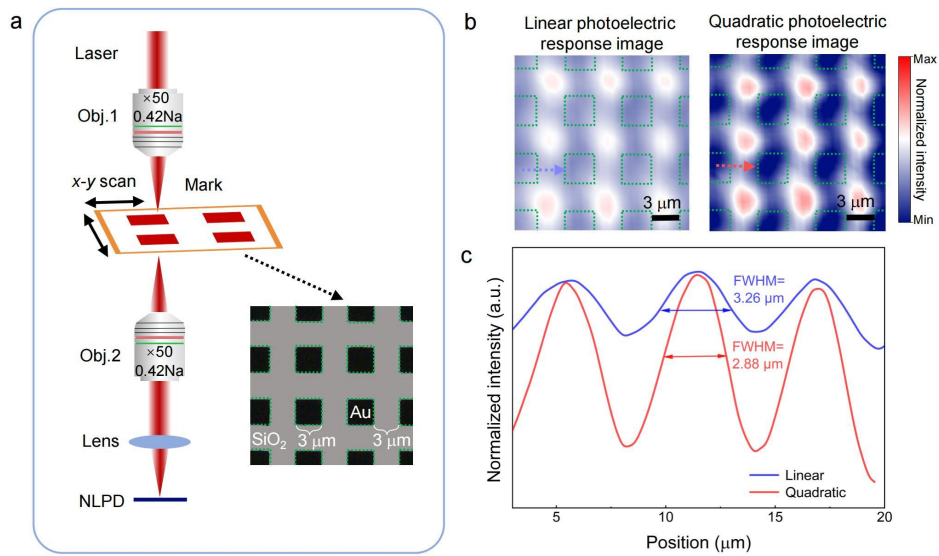

**Figure 5.** Improving spatial imaging resolution by the quadratic photoelectric conversion response of the NLPD than the linear photoelectric conversion response. **a** Experimental setup

for imaging an Au patterned mask. **b** Imaging results of the Au mask by the NLPD with the regimes of linear and quadratic photoelectric conversion responses. **c** Comparison of the FWHM of gray curves extracted from **b**.

## 3. Conclusion

In conclusion, we have demonstrated a NLPD based on a few-layer InSe *p-n* homojunction with a deterministic quadratic function between the photocurrent and the optical power. Thanks to the thickness of few atomic layers, the InSe flake could be easily doped into *p*- or *n*-type with electrostatic fields. By stacking the InSe few-layer on a split dual-gate, high-quality reconfigurable *p-n*, *n-p*, *p-p*, *n-n* homojunctions are realized with rectification ratios exceeding $10^6$. With the built-in electric field across the InSe *p-n* homojunctions, photocarriers in InSe generated from the light absorption could be separated and give rise to considerable photocurrent. Though the InSe *p-n* homojunction could perform as a photovoltaic photodetector with a high responsivity of 0.4 A/W for the light at 532 nm, we propose to operate it in a nonlinear photoelectric conversion mode. The employed few-layer InSe has significantly strong SHG, which could upconvert two photons with low energy into one photon with doubled energy. For light with low photon energy that can not be linearly absorbed by InSe, it could be absorbed by InSe after the SHG process if its doubled photon energy is larger than the bandgap of InSe. Consequently, the photocarriers are induced in the few-layer InSe, which generate photocurrent after their effective separations driven by the *p-n* homojunction. The combination of the SHG and photoelectric conversion gives rise to the NLPD. For the incident pulsed laser at the wavelength of 1550 nm, remarkable photocurrents are realized in the NLPD, presenting a quadratic dependence on the optical power. It is in great contrast to the performance of the InSe photodetector in sensing light at 532 nm, which has a linear function between photocurrent and optical power. The photoelectric response wavelength range of the NLPD is extended to 1750 nm, which is far beyond the wavelength range of 950 nm defined by InSe's bandgap. A high responsivity up to 0.534 A/W$^2$ is achieved, which is more than two orders of magnitude higher than that obtained in our previous work based on a GaSe/InSe vdWs heterojunction, and also surpass the responsivity of numerous NLPD utilizing alternative photoelectric conversion mechanisms.

The nonlinear dependence of photocurrents on the optical powers makes the NLPD have high sensitivity to the variation of optical intensity, which therefore enables high spatial resolution in measuring optical profiles. As examples, we first demonstrate the precise determination of the focused laser beam waist using the NLPD with a highly localized focal spot. Second, the NLPD is employed to implement imaging with an improved resolution compared with that of the linear photodetector. While the responsivity of the NLPD is still limited by the thickness of the InSe. In the future, by integrating the InSe *p-n* homojunction with optical resonators or waveguides to effectively enhance the light-material interaction,[41-44] the performance of the NLPD could be improved greatly. In addition to the improved performance in spatial measurements of optical intensity, relying on the combinations of all-optical mixing and photoelectric conversion, the proposed NLPD could also find applications in optical sampling, nonlinear signal processing, and optoelectronic frequency mixing.[11]

## 4. Methods

*Device Fabrication:* A pair of local bottom gate electrodes separated by 150 nm was fabricated using electron-beam lithography on a 300 nm thick $SiO_2$ film on a Si substrate. Few-layer InSe flakes were mechanically exfoliated with scotch tape from an ε-type InSe bulk crystal grown by the Bridgman method. Few-layer h-BN flakes were exfoliated from commercially available bulk crystals on PDMS (HQ Graphene Corporation). Afterwards, h-BN and InSe flakes were sequentially transferred to the bottom gates utilizing the dry transfer method. The Au (with the thickness of 90 nm) electrodes were prepared with electron beam evaporation on a Si substrate using a shadow mask. The Au electrodes were then picked up from the substrate with PDMS and transferred onto InSe layer as the source-drain electrodes. All the transfer processes were performed in the glove box immediately after the material exfoliation to prevent contamination of the interface. Subsequently, the fabricated devices are then annealed at 200 °C with Ar (including 5% $H_2$) protection for 2 hours to release stress and improve electrical contact.

*Characterizations*: The thicknesses of few-layer InSe flakes were verified by atomic force microscopy (Bruker Dimension icon). Raman and PL spectra of InSe flakes were collected by using a confocal micro-Raman system (Alpha300R, WITec) excited by 532 nm laser (spot size 400 nm, laser power 1.0 mW, resolution 0.02 $cm^{-1}$). The SHG measurement of InSe homojunction was carried out on a custom-built multiphoton nonlinear confocal optical microscope system. An optical parametric oscillator (Chromacity Inc.) was employed as the pulsed laser, which outputs picosecond laser pulses (with pulsed width of 4 ps and repetition rate of 100 MHz) at the tunable wavelength range between 1500 nm and 1750 nm. Then the laser beam was reflected by a dichroic mirror and focused on the sample through a 50× objective lens with a numerical aperture of 0.75. The generated SHG radiation was then back collected by the same objective lens and passes through the dichroic mirror, which is subsequently coupled to a spectrometer mounted with a cooled silicon CCD camera for spectra measurements. All the electrical/optoelectronic measurements were carried out by a semiconductor parameter analyzer (PDA FS380 Pro, Platform Design Automation) in a custom-built shielding box with continuous nitrogen flow to prevent degradation of the device performance during test. The 532 nm and 1550 nm laser beams were focused by the objective lens. For the photoelectric response speed, the modulation of the laser was realized using a mechanical chopper (Stanford SR540). For the photocurrent mapping, a *x-y-z* stage (Newport XYZ100) was used to scan in the *x-y* plane with the laser power and the focal plane was kept unchanged. While for the photocurrent scanning in the vertical direction, the stage was moved along the *z* axis and other conditions were kept unchanged.

**Supporting Information**

Supporting Information is available from the Wiley Online Library or from the author.

**Acknowledgements**

This project was primarily supported by the Key Research and Development Program (2022YFA1404800), National Natural Science Foundation of China (12374359, 62375225, 62104198, 62305270), Shaanxi Fundamental Science Research Project for Mathematics and Physics (22JSY004), Xi'an Science and Technology Plan Project (2023JH-ZCGJ-0023). The authors also thank the Analytical & Testing Center of NPU for their assistance in device fabrication and characterizations.

**Conflict of Interest**

The authors declare no conflict of interest.

**Supporting Information**

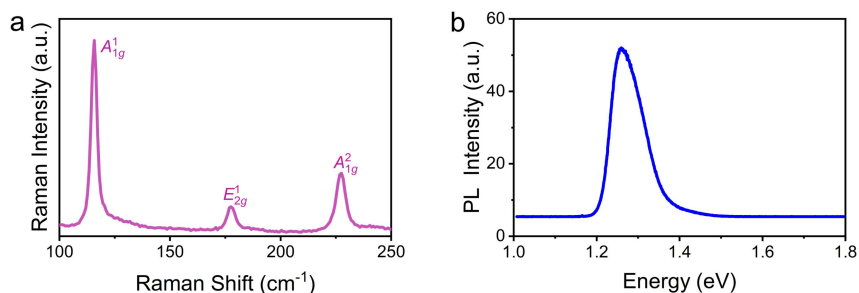

**Figure S1. a** Raman spectrum, **b** Photoluminescence spectrum of the employed InSe flake.

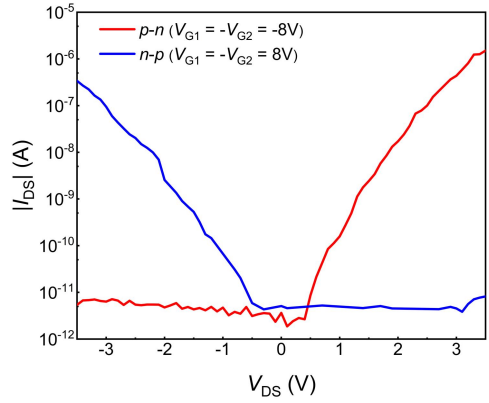

**Figure S2.** Output characteristics ($|I_{DS}|$-$V_{DS}$ curves) for *p-n* (red line) and *n-p* (blue line) homojunctions plotted on a semi-logarithmic scale.

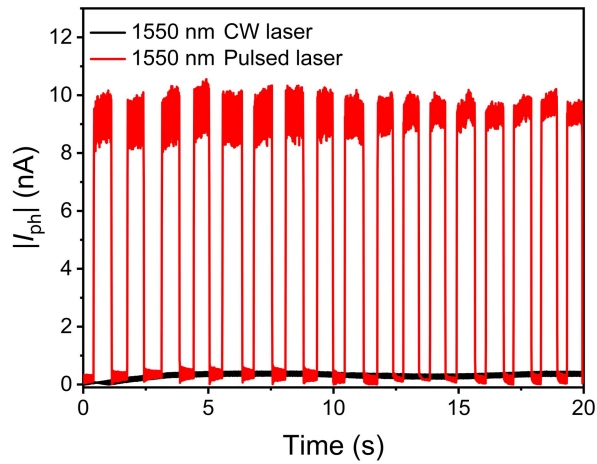

**Figure S3.** Photocurrent of the InSe *p-n* homojunction under the illuminations of a CW laser and a pulsed laser at the wavelength of 1550 nm, which have the same averaged power.

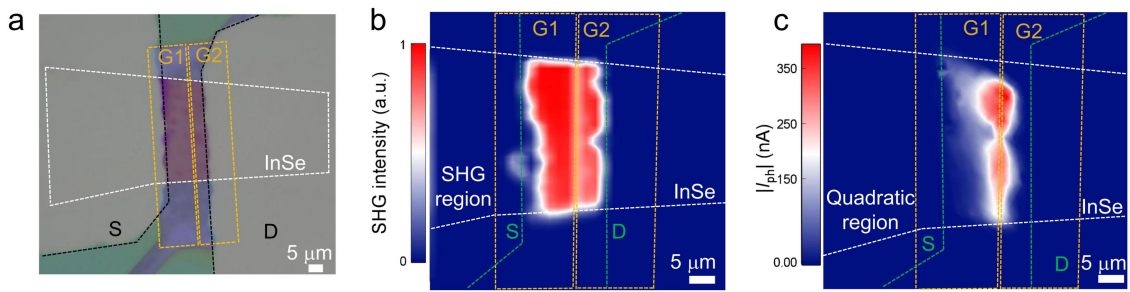

**Figure S4.** Spatial mappings of SHG and photocurrents with the illumination of a 1550 nm pulsed laser. **a** Optical micrograph of the InSe homojunction device. **b** SHG mapping of the device. SHG signals are uniformly distributed throughout the whole InSe channel. **c** Quadratic nonlinear photocurrent mapping of the device. The photocurrent predominantly distributes around the gap region between the two back gate electrodes.

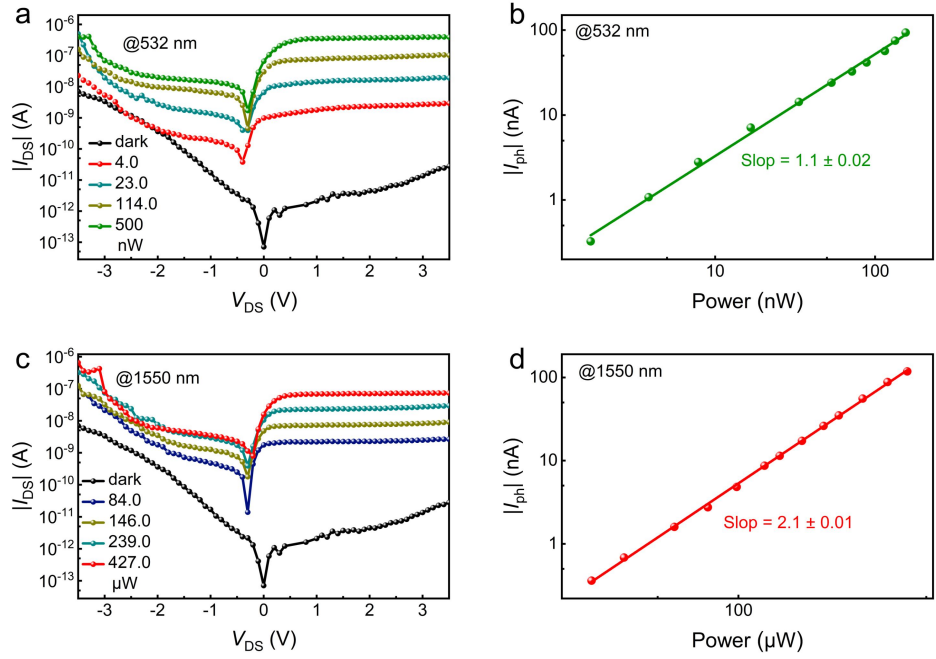

**Figure S5.** Linear and quadratic nonlinear photoelectric response of the NLPD based on the InSe *n-p* homojunction ($V_{G1}$ = 8V, $V_{G2}$ = -8V). **a** $|I_{DS}|$-$V_{DS}$ curves obtained with the illumination of a 532 nm continuous wave laser with photon energy larger than InSe bandgap under different optical powers. **b** Laser power dependence of the photocurrent at $V_{DS}$ = 3.5 V for the illumination of the 532 nm laser, presenting a slope of 1.1. **c** $|I_{DS}|$-$V_{DS}$ curves obtained with the illumination of a 1550 nm pulsed laser with photon energy smaller than InSe bandgap under different optical powers. **d** Laser power dependence of the photocurrent at $V_{DS}$ = 3.5 V for the illumination of the 1550 nm laser, presenting a slope of 2.1.